\documentclass[12pt]{article}

\usepackage{epsfig}

\begin{document}

\newcommand{\nc}{\newcommand}
\nc{\bc}{\begin{center}}
\nc{\ec}{\end{center}}
\nc{\be}{\begin{equation}}
\nc{\ee}{\end{equation}}
\nc{\bfg}{\begin{figure}}
\nc{\efg}{\end{figure}}
\nc{\bbib}[1]{}

\title{ \bf Kaluza-Klein gravitons at the LHC and in extensive air showers%
\thanks{Talk given at the {\em International Conference on Interconnection
between High Energy Physics and Astroparticle Physics: From
Colliders to Cosmic Rays}, 7-13 September 2005, Prague, Czech
Republic.} }

\author{A.V. Kisselev and V.A. Petrov \\
\small Institute for High Energy Physics, 142281 Protvino, Russia}

\date{}

\maketitle

\thispagestyle{empty}

\begin{abstract}
The small curvature option of the Randall-Sundrum model with two
branes is considered which has almost continuous spectrum of
low-mass Kaluza-Klein gravitons. It is shown that gravity effects
related with these excitations can be detected in double
diffractive events at the LHC and in inclined air showers induced
by interactions of cosmic neutrinos with atmospheric nucleons at
ultra-high energies.
\end{abstract}

\section*{Introduction}

One of the most important problem of the particle physics is the
problem of hierarchy between the electro-weak and Planck scales.
To explain this hierarchy, a number of theories with extra spacial
dimensions have been proposed (see, for instance,
Ref.~\cite{ADD}). The model which solves the problem most
economically is the so-called RS1 model with a single extra
dimension~\cite{RS1}. The background (warped) metric of the model
is of the form:
\be\label{metric}
ds^2 = e^{2 \kappa (\pi r_c - |y|)} \, \eta_{\mu \nu} \, dx^{\mu}
dx^{\nu} + dy^2 \;.
\ee
Here $y = r_c \, \theta$ ($-\pi \leq \theta \leq \pi$), $r_c$
being a ``radius'' of the extra dimension, while $\{x^{\mu}\}$,
$\mu = 0,1,2,3$, are the coordinates in four-dimensional
space-time. The parameter $\kappa$ defines the scalar curvature in
five dimensions. Note that the points $(x^{\mu}, y)$ and
$(x^{\mu}, -y)$ are identified, and the periodicity condition,
$(x^{\mu}, y) = (x_{\mu}, y + 2 \pi r_c)$, is imposed. The tensor
$\eta_{\mu \nu}$ in~(\ref{metric}) is the Minkowski metric.

There are two 3-dimensional branes with equal and opposite
tensions located at the point $y = 0$ (called the \emph{Plank
brane}) and point $y =  \pi r_c$ (referred to as the \emph{TeV
brane}). All SM fields are constrained to the TeV brane. Then one
can derive the relation between the 4-dimensional (reduced) Planck
scale, $\bar{M}_{Pl}$, and (reduced) gravity scale in five
dimensions, $\bar{M}_5$,
\be\label{hierarchy}
\bar{M}_{Pl}^2 = \frac{\bar{M}_5^3}{\kappa} \left(e^{2 \pi \kappa
r_c} - 1 \right) \;.
\ee

The masses of the Kaluza-Klein (KK) graviton excitations are given
by
\be\label{KK_masses}
m_n = x_n \, \kappa , \qquad n=1,2 \ldots
\;,
\ee
where $x_n$ are zeros of the Bessel function $J_1(x)$. The zero
graviton mode, $h^{(0)}_{\mu \nu}$, and massive graviton modes,
$h^{(n)}_{\mu \nu}$, are coupled to the energy-momentum tensor of
the matter, $T^{\mu \nu}$, as follows:
\be\label{Lagrangian}
\mathcal{L}_{int} = - \frac{1}{\bar{M}_{Pl}} \, T^{\mu \nu} \,
h^{(0)}_{\mu \nu} - \frac{1}{\Lambda_{\pi}} \, T^{\mu \nu} \,
\sum_{n=1}^{\infty} h^{(n)}_{\mu \nu} \;,
\ee
with
\be\label{Lambda}
\Lambda_{\pi} = \left( \frac{\bar{M}_5^{3}}{\kappa} \right)^{\!
1/2}
\ee
being a physical scale on the TeV brane.

In Ref.~\cite{small_curv} it has been proposed to study the ``{\em
small curvature option}'' of the RS1 model:
\be\label{small_curv}
\kappa \ll \bar{M}_5 \sim 1 \mathrm{\ TeV} \;.
\ee
In such a case, we get an almost continuous spectrum of low-mass
graviton excitations with a small mass splitting $\Delta m \simeq
\pi \kappa$. Note that in the usually used scenario of the RS1
model one has a series of KK graviton resonances with the lightest
one having a mass of order 1 TeV.

\section{KK graviton production at the LHC}

One of the possibilities to check the RS1 model with the small
curvature~(\ref{small_curv}) is to search for an exclusive production of the
KK gravitons in double diffractive events at the LHC.

Since the widths of the graviton with the KK number $n$ and mass
$m_n$ is extremely small,
\be\label{width}
\Gamma_n \simeq 0.1 \, \frac{m_n^3}{\Lambda_{\pi}^2} \; ,
\ee
a distinctive signature of the production of such fields will be
an imbalance in missing mass of final states. Because of the small
mass splitting in the KK graviton spectrum, we expect a continuous
distribution in missing mass.

In other words, one has to look for the process
\be\label{EDDE}
p + p \rightarrow p + \mbox{``nothing''} + p \; .
\ee

The distribution in missing mass $M_{miss}$ for the production of
the KK gravitons in double diffractive events was calculated in
Ref.~\cite{EDDE}. The results of these calculation are presented
in Fig.~\ref{fig:missing_mass} for various values of $\bar{M}_5$,
the gravity scale in five warped dimensions.

\bfg[h]
\bc
\epsfysize=16.5em \epsffile{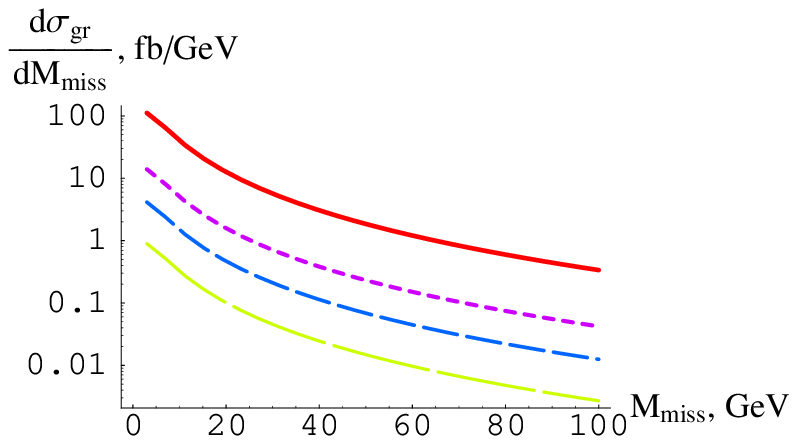}
\ec
\vspace{-5mm}
\caption{The distribution in the missing mass in the double
diffractive production of the KK gravitons. The curves correspond
(from top to bottom) to $\bar{M}_5 = $ 1~TeV, 2~TeV, 3~TeV, and
5~TeV.}
\label{fig:missing_mass}
\efg

Let us stress that $d\sigma_{gr}/dM_{miss}$ is defined by
$\bar{M}_5$ only, not by the values of $\kappa$ and
$\Lambda_{\pi}$ separately. The smallness of the graviton coupling
($\sim 1/\Lambda_{\pi}^2$) is compensated by the large number of
the produced gravitons ($\sim 1/\kappa$). As a result, the
corresponding cross sections appeared to be large enough, as one
can see in the next Fig.~\ref{fig:graviton_prod}.

\bfg[h]
\bc
\epsfysize=16.5em \epsffile{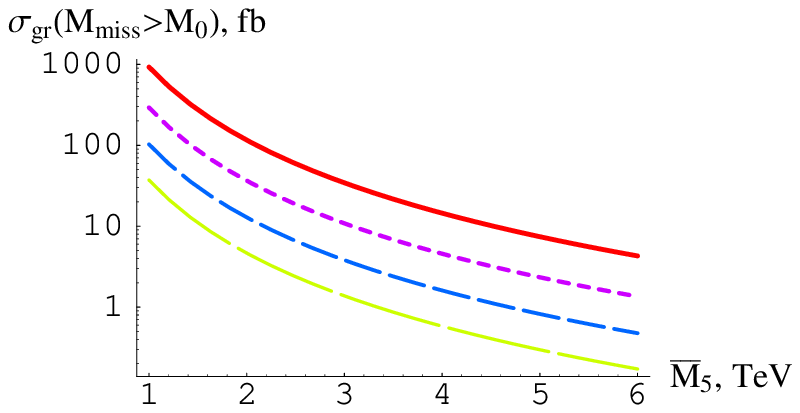}
\ec
\vspace{-5mm}
\caption{The cross section for double diffractive production of
the KK gravitons with masses larger that $M_0$ as a function of
$\bar{M}_5$. The curves correspond (from top to bottom) to $M_0 =
$ 3~GeV, 14~GeV, 30~GeV, and 50~GeV. The lowest value of $M_0$ is
imposed to avoid the influence of the soft photon production, the
highest concerns a neutrino background from $Z$-decays.}
\label{fig:graviton_prod}
\efg

We expect that the signals from the production of the KK gravitons
in double diffractive events could be detected at the LHC in a
joint experiment of the CMS and TOTEM Collaborations~\cite{TOTEM}.

\section{Gravi-Reggeons and cosmic neutrinos}

Another possibility to detect effects induced by low-mass KK
gravitons is to look for their contributions to the scattering of
the brane fields in the trans-Planckian kinematical
region~\cite{small_curv}
\be\label{kinem_region}
\sqrt{s} \gg \bar{M}_5, \qquad s \gg -t
\;,
\ee
with $\sqrt{s}$ being the colliding energy and $t = - q_{\bot}^2$
being the four-dimensional momentum transfer. Note that
regime~(\ref{kinem_region}) dominates the whole value of both
elastic and inelastic cross sections.

In the eikonal approximation, an elastic scattering amplitude in
the kinematical region~(\ref{kinem_region}) is given by the sum of
gravi-Reggeons, i.e. reggeized gravitons in $t$-channel. Because
of a presence of extra dimension, the Regge trajectory of the
graviton is splitting into an infinite sequence of trajectories
enumerated by the KK number $n$~\cite{gravi-Reggeons}:
\be\label{trajectories}
\alpha_n(t) = 2 + \alpha_g' t  - \alpha_g' \, m_n^2,
\quad n = 0, 1, \ldots.
\ee

Let us now consider scattering of ultra-high energy cosmic
neutrinos off the atmospheric nucleons in order to compare effects
induced by extra dimension with the SM predictions~\cite{SM}. The
gravity contribution to the neutrino-proton cross section
\be\label{cross_sec_had}
\sigma^{\nu \rm p}_{\rm in}(s) = \int \! d^2 b \left\{ 1 - \exp
\big[- 2 \mbox{\rm Im} \, \chi_{\nu \rm p}(s,b) \big] \right\} \;,
\ee
is defined by the eikonal
\be\label{eikonal_had}
\chi_{\nu \rm p}(s,b) = \frac{1}{4\pi s} \int\limits_0^{\infty}
\! q_{\bot} d q_{\bot} \, J_0(q_{\bot} b) \, A_{\nu \rm p}^B(s,
-q_{\bot}^2) \;,
\ee
where $b$ denotes an impact parameter.

The Born amplitude was calculated in Ref.~\cite{small_curv}:
\be\label{Born_ampl}
A_{\nu \rm p}^B(s, t) = \frac{\alpha'_g \, s^2}{2 \sqrt{\pi} \,
\bar{M}_5^3} \, \sum_i \int \! dx \, x^2 \, \frac{1}{ R_g
(sx)} \, \exp \left[ t \, R_g^2(sx) \right] F_i(x,t) \; ,
\ee
where $F_i(x,t)$ is a $t$-dependent distribution of parton $i$ in
momentum fraction $x$ inside the proton (see \cite{small_curv} for
details). It coincides with a starndard parton distribution,
$f_i(x)$, at $t=0$. The quantity $ R_g(s) = \alpha'_g \ln s$ is
the gravitational interaction radius, where $\alpha'_g$ is the
gravi-Reggeon slope~(\ref{trajectories}).

The gravitational part of the inelastic cross section is presented
in Fig.~\ref{fig:sigma_had} in comparison with the SM prediction,
$\sigma_{\rm SM}$, and black hole production cross section,
$\sigma_{\rm bh}$. For the latter, a geometrical form,
$\sigma_{\rm bh} = \pi R_S^2(s)$, is assumed, with 5-di\-mensional
Schwarzschild radius~\cite{Schw_radius}
\be\label{Schw_radius}
R_S(s) = \frac{1}{\sqrt{3}\pi} \, \frac{s^{1/4}}{\bar{M}_5^{3/2}}
\;.
\ee

\bfg[h]
\bc
\epsfysize=16.5em \epsffile{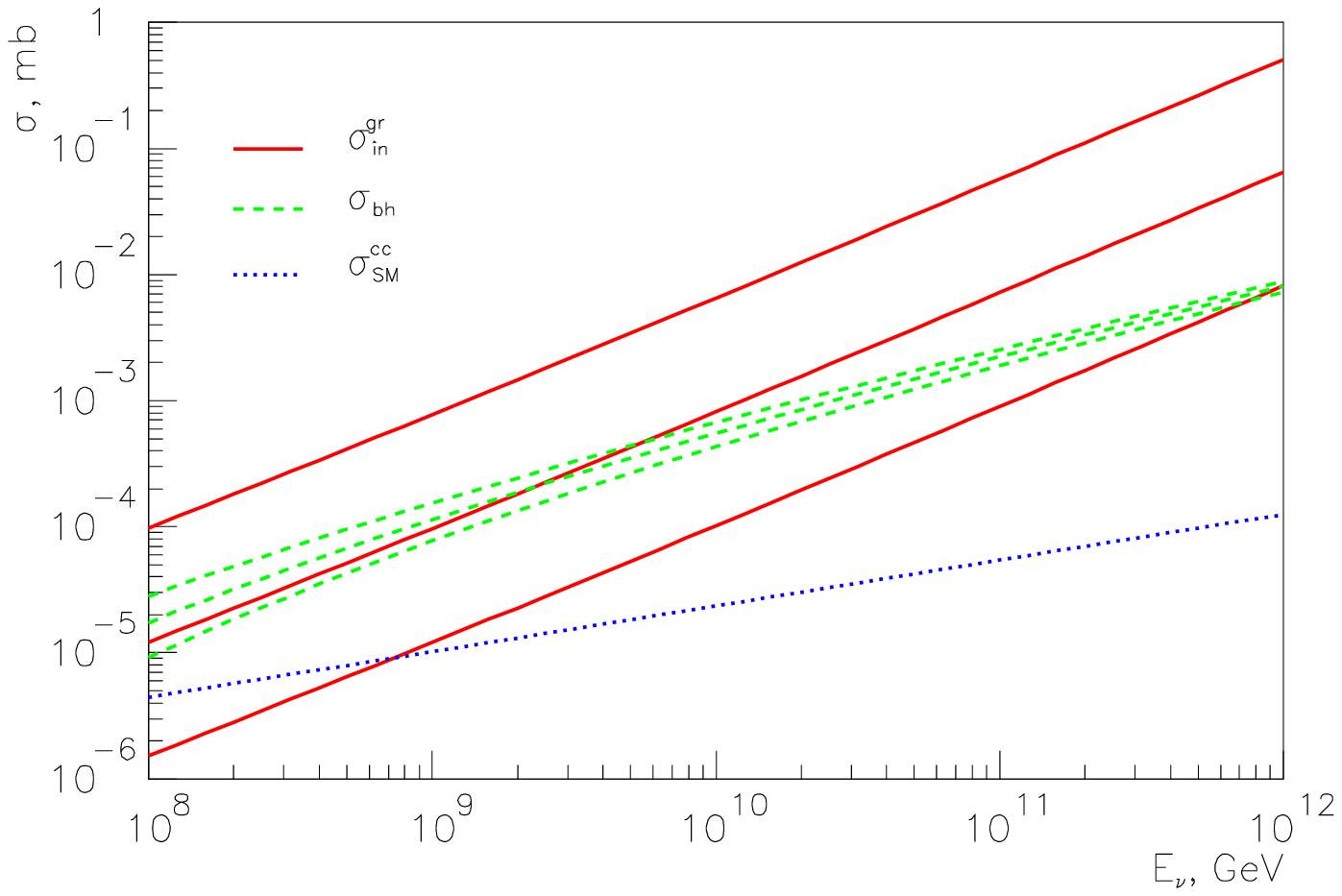}
\ec
\vspace{-5mm}
\caption{The gravitational inelastic neutrino-proton
cross-sections (solid lines) vs. black hole production cross
sections (dashed lines) and  SM cross section (dotted line). The
solid curves correspond to $\bar{M}_5 = 0.25$ TeV, 0.5 TeV, 1 TeV
(from the top). The dash lines correspond to $\bar{M}_5 = 0.5$ TeV
and $M_{\rm bh}^{\rm min} = 0.5$ TeV, 1~TeV, 2 TeV (from the
top).}
\label{fig:sigma_had}
\efg
As one can see, gravi-Reggeon interactions can dominate black hole
production at $E_{\nu} > 10^9-10^{10}$ GeV, depending on the
gravity scale $\bar{M}_5$ and minimal value of the black hole mass
$M_{\rm bh}^{\rm min}$.

Let us stress that usually the eikonalization is made at the
parton level, e.i. before the convolution of the cross section
with the parton distributions. In such a case, the neutrino-proton
cross section is of the form:
\be\label{cross_sec_par}
\sigma_{\rm in}^{\nu \rm p} (s) = \sum_i \int \! dx \,
\hat{\sigma}(xs) \, f_i(x) \;,
\ee
where $\hat{\sigma}_i$ is the cross section for the scattering of
the neutrino off parton $i$. For the gravitational interaction, it
does not depend on $i$ and it is equal to
\be\label{cross_sec_subproc}
\hat{\sigma}(s) = \int \! d^2 b \left\{ 1 - \exp \big[- 2
\mbox{\rm Im} \, \hat{\chi}(s,b) \big] \right\} \;.
\ee

The gravitational part of the eikonal for the neutrino scattering
off a parton looks like~\cite{small_curv}
\be\label{eikonal_par}
\mbox{\rm Im} \, \hat{\chi}(s,b) = \frac{\alpha'_g \, s}{16
\sqrt{\pi} \, R_g^3(s) \, \bar{M}_5^3} \, \exp \Big[ -b^2/4 R_g^2
(s)\Big] \;.
\ee

The results of numerical calculations made with the use of
formulas~(\ref{cross_sec_par})-(\ref{eikonal_par}) are presented
in Fig.~\ref{fig:sigma_par}. It shows that the inelastic cross
sections rise with energy slower that in the case when the
eikonalization is made at the hadronic level (see
Eqs.~(\ref{cross_sec_had})-(\ref{Born_ampl})).

\bfg[h]
\bc
\epsfysize=16.5em \epsffile{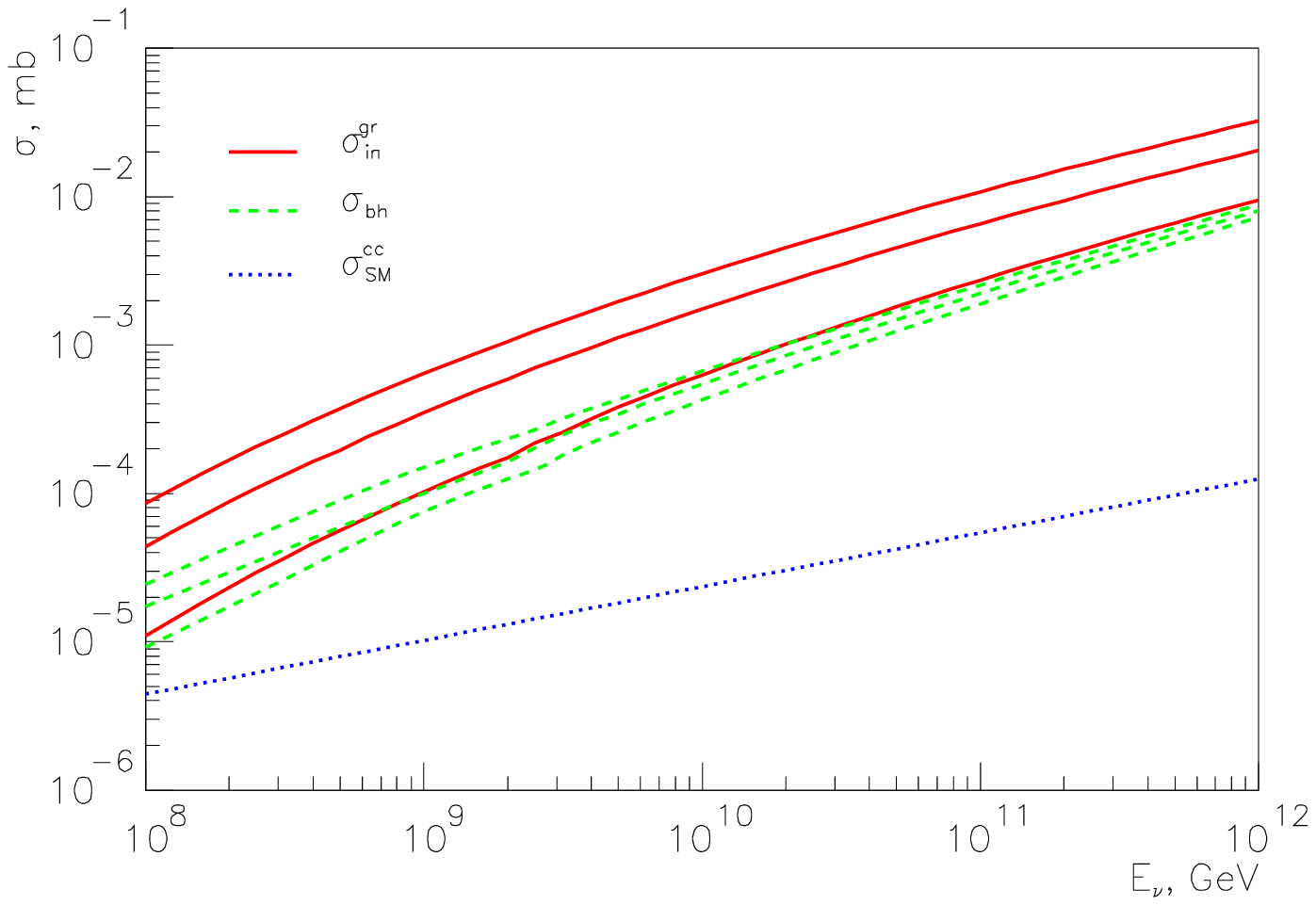}
\ec
\vspace{-5mm}
\caption{The same as in Fig.~\ref{fig:sigma_had}, but with
the eikonalization made at the partonic level (see the main text
for details).}
\label{fig:sigma_par}
\efg

These neutrino-proton cross sections can be probed by the Pierre
Auger Observatory~\cite{AUGER}. In order to isolate
neutrino-induced events, inclined (quasi-horizontal) air showers
should be looked for~\cite{incl_showers}. For the
``Waxman-Bahcall'' neutrino flux~\cite{WB}, we expect the
following number of the inclined air showers induced by
ultra-high energy cosmic neutrinos (with zenith angle $\theta >
70^{\circ}$):
\be\label{event_rate}
N_{\rm ev} =
   \cases{4.9 \mbox{ yr}^{-1}, & $\bar{M}_5 = 1$ TeV \cr
          1.6 \mbox{ yr}^{-1}, & $\bar{M}_5 = 2$ TeV \cr}
\ee
These estimates should be compared with the SM prediction, 0.08
events per year for the same neutrino flux.

\end{document}